\documentclass{WileyMSP-template}
\usepackage{amsmath}
\begin{document}

\pagestyle{fancy}
\rhead{\includegraphics[width=2.5cm]{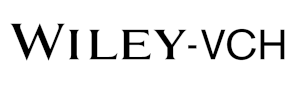}}

\title{Ultrashort Pulse Laser Annealing of Amorphous Atomic Layer Deposited MoS\textsubscript{2} Films}

\maketitle


\author{Malte J. M. J. Becher*}
\author{Julia Jagosz}
\author{Rahel-Manuela Neubieser}
\author{Jan-Lucas Wree}
\author{Anjana Devi}
\author{Marvin Michel}
\author{Claudia Bock}
\author{Evgeny L. Gurevich}
\author{Andreas Ostendorf}

\begin{affiliations}
M. J. M. J. Becher, E. L. Gurevich , A. Ostendorf\\
Applied Laser Technologies\\
Ruhr University Bochum\\
Universitätsstr. 150\\
44801 Bochum, Germany\\
Email Address: malte.becher@rub.de

R.-M. Neubieser, M. Michel, A. Devi\\
Fraunhofer IMS\\
Finkenstraße 61\\
47057 Duisburg, Germany

J.-L. Wree, A. Devi\\
Inorganic Materials Chemistry\\
Ruhr University Bochum\\
Universitätsstr. 150\\
44801 Bochum, Germany

J. Jagosz, C.Bock\\
Microsystems Technology\\
Ruhr University Bochum\\
Universitätsstr. 150\\
44801 Bochum, Germany\\

E. L. Gurevich\\
Laser Center FH Münster\\
Stegerwaldstr. 39\\
48565 Steinfurt, Germany

\end{affiliations}



\begin{abstract}

Thin films of molybendum disulfide grown via thermal atomic layer deposition at low temperatures, suitable for temperature sensible substrates, can be amorphous. To avoid a high temperature post treatment of the whole sample, which can cause thermal degradation of the substrate or other layers, a ultrashort pulse (usp) laser-induced transformation to crystalline layers is one of the most promising routes. In this paper we report the crystallization of amorphous MoS\textsubscript{2} layers processed with ultrashort laser pulses. The amorphous MoS\textsubscript{2} films were deposited by atomic layer deposition (ALD) and exposed to picosecond laser pulses ($\lambda$ = 532 nm). The crystallization and the influence of the processing parameters on the film morphology were analyzed in detail by Raman spectroscopy and scanning electron microscopy. Furthermore, a transition of amorphous MoS\textsubscript{2} by laser annealing to self-organized patterns is demonstrated and a possible process mechanism for the ultrashort pulse laser annealing is discussed. Finally, the usp laser annealed films were compared to thermally and continuous wave (cw) laser annealed samples.

\end{abstract}


\section{Introduction}

In the next few years the demand for flexible electronics in medicine, sensors, communication and wearables will grow strongly \cite{Gao.2019, Jiang.2021, Kim.2011, Sun.2007}. Promising materials for these electronic devices are 2D transition metal dichalcogenides (TMDCs), with molybdenum disulfide (MoS\textsubscript{2}) as the most researched representative, because of their unique electronic, mechanical and optical properties \cite{Wang.2012,Jariwala.2014}. For industrial usage, the generation of large-area homogeneous MoS\textsubscript{2} films is necessary. Typical deposition methods like chemical vapor deposition (CVD) require high deposition temperatures to produce large-area crystalline MoS\textsubscript{2} films. These temperatures disable the use of low melting point substrates like glass or polymers required for flexible electronics. ALD processes uses deposition temperatures ranging from 100 °C to 250 °C, producing large area films of MoS\textsubscript{2}. These films feature amorphous \cite{Neubieser.2022} or poly-crystalline \cite{Mattinen.2022} properties. For applications in high efficiency electronics, like thin-film transistors, a mono-crystalline film is desirable, which typically leads to a post treatment of films, which do not fulfill the requirements. The classic approach to improve the crystallinity in MoS\textsubscript{2} films is thermal annealing in a furnace \cite{Tan.2014,Wong.2017,Johari.2021,Kim.2018b}.  Tan et al. \cite{Tan.2014} and Wong et al. \cite{Wong.2017} have shown that a thermal annealing in a sulfur or nitrogen atmosphere with temperatures from 450 °C to 1000 °C increases the crystallinity of amorphous MoS\textsubscript{2} films. This temperature range is not suitable for flexible substrates like Kapton/Polyimide (PI) or polyethylene terephthalate (PET). Due to the temperature sensibility of flexible substrates other more localized methods for the annealing of MoS\textsubscript{2} films need to be explored. 
Laser processing of materials can modify the crystal structure to achieve a higher degree of crystallinity. Amorphous silicon films and optical fibers were crystallized by long-pulsed \cite{Coucheron.2016} or continuous wave \cite{Sasaki.2020, Healy.2014} laser. Adopting this method to MoS\textsubscript{2}, the crystallinity of amorphous magnetron sputtered MoS\textsubscript{2} films on polymer was increased \cite{Sirota.2019, McConney.2016}. The heating by relatively  long laser pulses gives the atoms enough time to rearrange themselves in the lattice. The use of shorter laser pulses (pico- or femtosecond pulses) is characterized by high cooling rates (up to 10\textsuperscript{12}~Ks$^{-1}$) \cite{Duff.2007} and hence, it is often used to disorder crystalline structures \cite{Liu.1979, GarciaLechuga.2021}. Nevertheless, usp laser-induced crystallization of silicon layers and titanium oxides \cite{Hoppius.2018, Zhan.2019} are reported employing a pulse-to-pulse temperature accumulation effect \cite{Gamaly.1999}. For amorphous MoS\textsubscript{2} a first laser-induced crystallization was reported, as a crystalline ring was formed during the processing with nanosecond Bessel - beam pulses\cite{Rai.2020}.
In this paper we demonstrated that amorphous MoS\textsubscript{2}, deposited by low-temperature ALD, can be two-dimensional crystallized upon irradiation with highly overlapped picosecond laser pulses and without any ablation. Moreover, we demonstrate formation of self-organized nanopatterns in the MoS\textsubscript{2} layer, as the laser power increases.

\begin{figure}
	\includegraphics[width=0.49\linewidth]{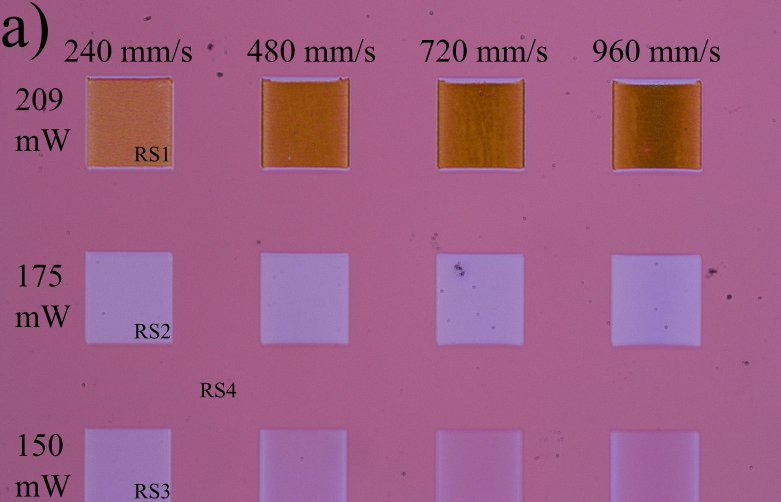}
		\includegraphics[width=0.49\linewidth]{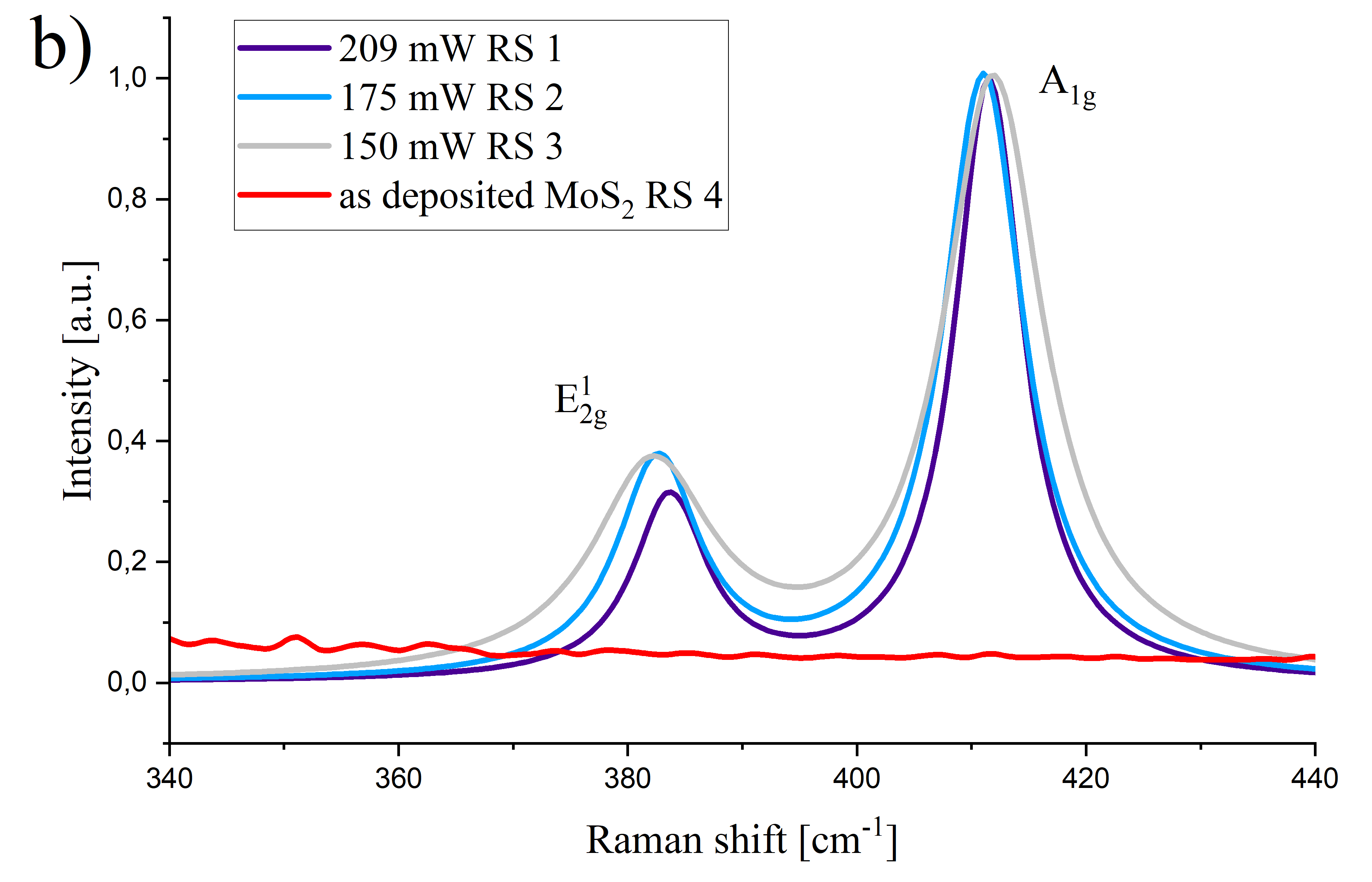}
	\caption{a) Optical micrograph image of a usp laser processed MoS\textsubscript{2} sample (thickness = 40 nm) prepared by ALD at $\theta$~=~100~\degree C. The laser power was varied from 100 mW to 209 mW in 25 mW steps.  b) Raman spectra of amorphous and usp laser annealed MoS\textsubscript{2} deposited by ALD (taken from the positions RS1-4). The laser annealed areas show the characteristic $\text{E}^1_{2g}$ and $\text{A}_{1g}$ peaks, whereas these peaks are missing in the pristine film.}\label{Proben_uebersicht}
\end{figure}

\section{Results}
\subsection{Characterization of usp laser processed MoS\textsubscript{2}}
For the crystallization experiments 250 µm x 250 µm squares were processed on the MoS\textsubscript{2} sample (thickness~=~40~nm on a SiO$_2$/Si thickness~=~200~nm substrate) with the usp laser pulses. The laser average powers used during the experiments varied from P\textsubscript{L,min}~=~100~mW to P\textsubscript{L,max}~=~209~mW with scanning speeds from v\textsubscript{scan,min}~=~240~mm~s$^{-1}$ to v\textsubscript{scan,max}~=~2640~mm~s$^{-1}$ with step~=~240~mm s$^{-1}$.  Figure \ref{Proben_uebersicht}(a) shows several areas processed at different parameters. The color change of the processed areas indicates a modification of the films in contrast to the pristine film. Laser powers P\textsubscript{L} $<$ 150 mW did not induce any noticeable color modification. The color change is more pronounced for lower scanning velocities, whereas for higher scanning speeds some inhomogeneity can be observed at the edge of the laser processed areas. The latter can be assigned to the acceleration and deceleration of the laser beam at the start and the end of the trajectories. A second color change becomes visible at P\textsubscript{L} $\geq$ 209 mW.

\begin{figure}
		\includegraphics[width=0.5\linewidth]{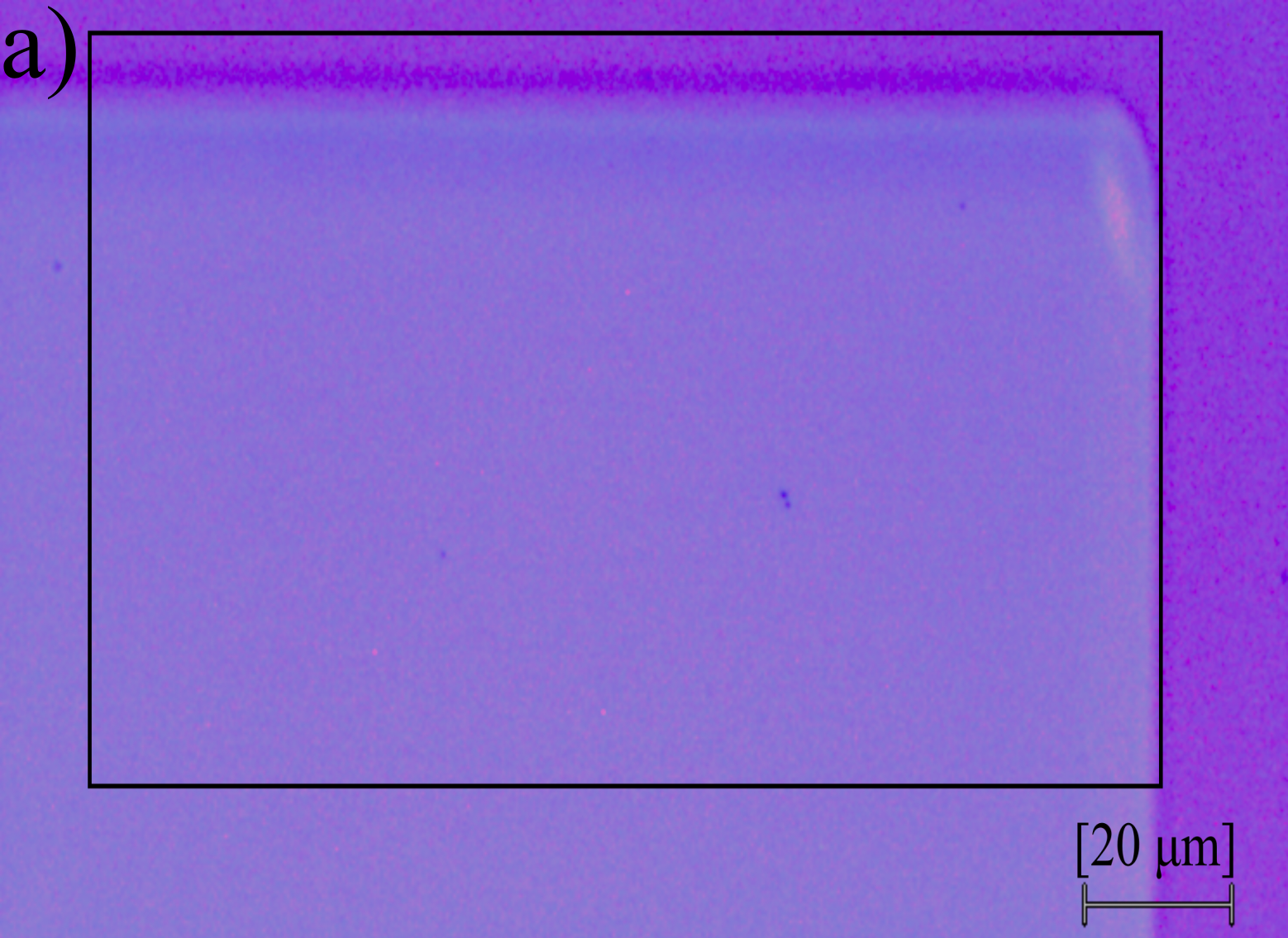}
		\includegraphics[width=0.5\linewidth]{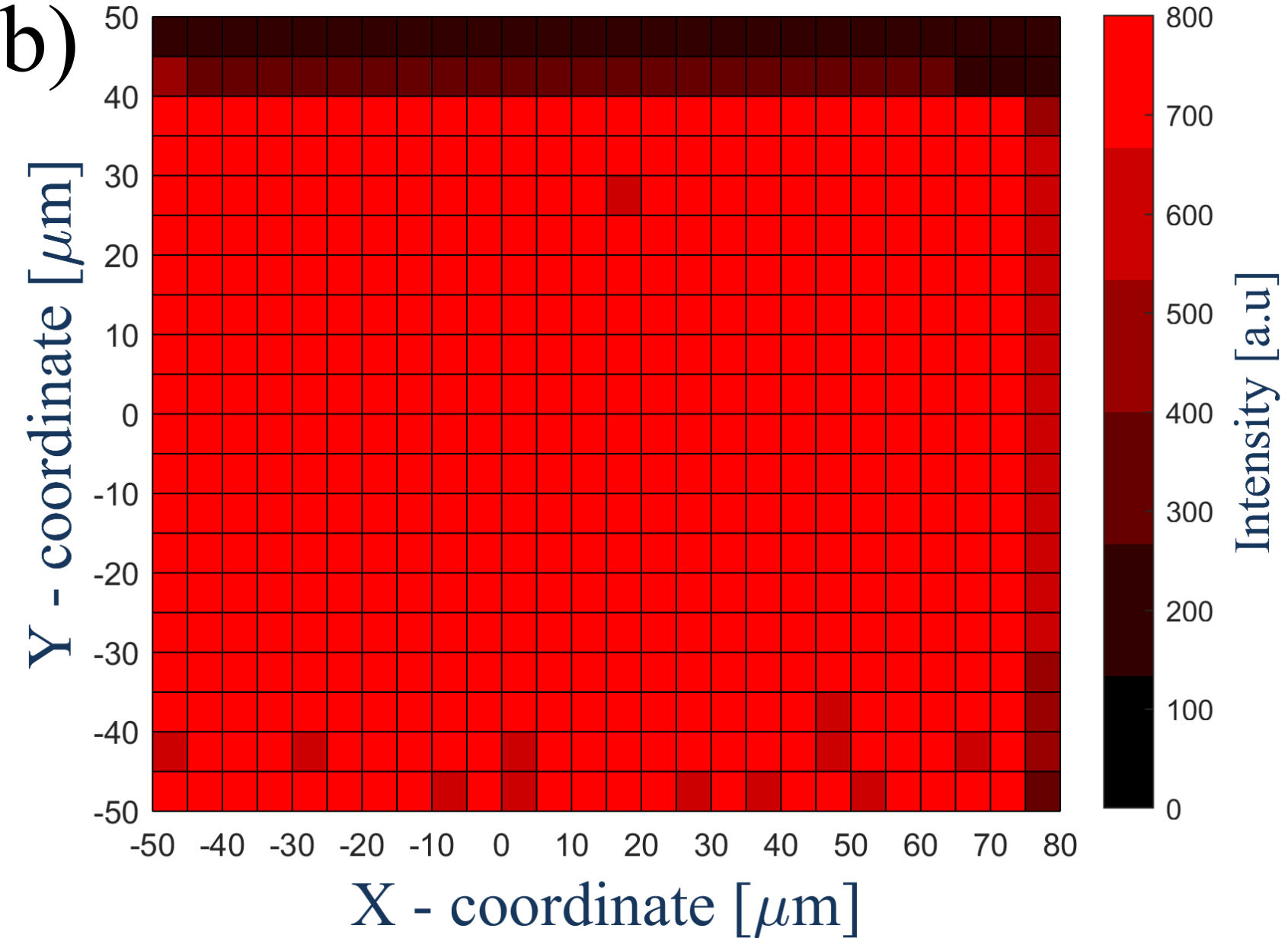}
	\caption{The color change of the processed area (P\textsubscript{L} = 175 mW, v\textsubscript{scan,min} = 240 mm s$^{-1}$) (bottom, left) in the light microscopy image a) indicates a modification of the original film (top, right). The Raman mapping b) of the $\text{A}_{1g}$ - peak (410~cm\textsuperscript{-1}) of the inset in a) demonstrates, that the laser annealing crystallized the processed area. The dark bar on top of the mapping represents the unprocessed area, where the intensity of the $\text{A}_{1g}$ - peak is absent.}\label{Raman_mapping}
\end{figure}

The evaluation of the crystallinity of the film was performed by Raman spectroscopy. Bulk crystalline MoS\textsubscript{2} has two characteristic Raman peaks that have their origin in the in - plane vibration ($\text{E}^1_{2g}$ - peak $\sim$ 383.0~cm\textsuperscript{-1}) and out-of-plane vibration ($\text{A}_{1g}$ - peak $\sim$ 407.8 cm\textsuperscript{-1}) of the Mo and S atoms \cite{Li.2012}. The diagram in Figure \ref{Proben_uebersicht}(b) shows the absence of the characteristic Raman peaks of amorphous, as deposited MoS\textsubscript{2}, allowing the detection of crystalline and amorphous MoS\textsubscript{2} films. A continuous, areal crystallization can be proven by Raman mappings, where local resolved information on a sample are gathered. Figure \ref{Raman_mapping}(a) shows the light microscopy image and (b) the Raman mapping of the A\textsubscript{1g} - peak of a MoS\textsubscript{2} sample processed with the usp laser (P\textsubscript{L} = 175 mW, v\textsubscript{scan} = 240 mm s$^{-1}$). The light microscopy image shows a color change in the film of MoS\textsubscript{2}, whereas the bottom, left side of the image, is the processed (crystalline) film and the top, right side is the original (amorphous) film. This behavior is also seen in the Raman mapping ($\text{A}_{1g}$ = 410 cm\textsuperscript{-1}), where the intensive red color imply a higher amplitude of the $\text{A}_{1g}$ - peak and darker colors implies weaker (or absence) $\text{A}_{1g}$ - peak intensities. The relative standard deviation of the pixels measured over the laser-exposed area in Figure \ref{Raman_mapping}(b) was 3.42 \%, demonstrating a high homogeneity of the laser annealing. 
Besides, the information about the successful crystallization, Raman spectra also provide additional information about the quality of the processed surface. Figure \ref{Proben_uebersicht}(b) shows the influence of different laser powers (v\textsubscript{scan} = 240 mm s$^{-1}$) on the Raman spectrum. For laser powers below P\textsubscript{L}~=~150~mW no crystalline peaks in the Raman spectrum were detected, which correlates with no optical modification of the film. The amplitudes of the spectra were normalized using the $\text{A}_{1g}$ - peak. The diagram shows that the peak full width at half maximum (FWHM) of both Raman peaks in each measured spectrum decreases with increasing laser power, starting from FWHM\textsubscript{$\text{E}^1_{2g}$}~=~12.8~cm\textsuperscript{-1} and FWHM\textsubscript{$\text{A}_{1g}$}~=~10.6~cm\textsuperscript{-1} at P\textsubscript{L}~=~150 mW, to FWHM\textsubscript{$\text{E}^1_{2g}$}~=~7.8 cm\textsuperscript{-1} and FWHM\textsubscript{$\text{A}_{1g}$}~=~7.2~cm\textsuperscript{-1} at P\textsubscript{L}~=~209~mW. This behavior can be explained by an increase in film crystallinity at higher laser powers \cite{Andrzejewski.2018}. Besides, the more narrow peaks, a small redshift of the $\text{E}^1_{2g}$ - peaks of the 150 mW and 175 mW spectra to the 209 mW spectra can be observed. 
\begin{figure}
			\includegraphics[width=0.49\linewidth]{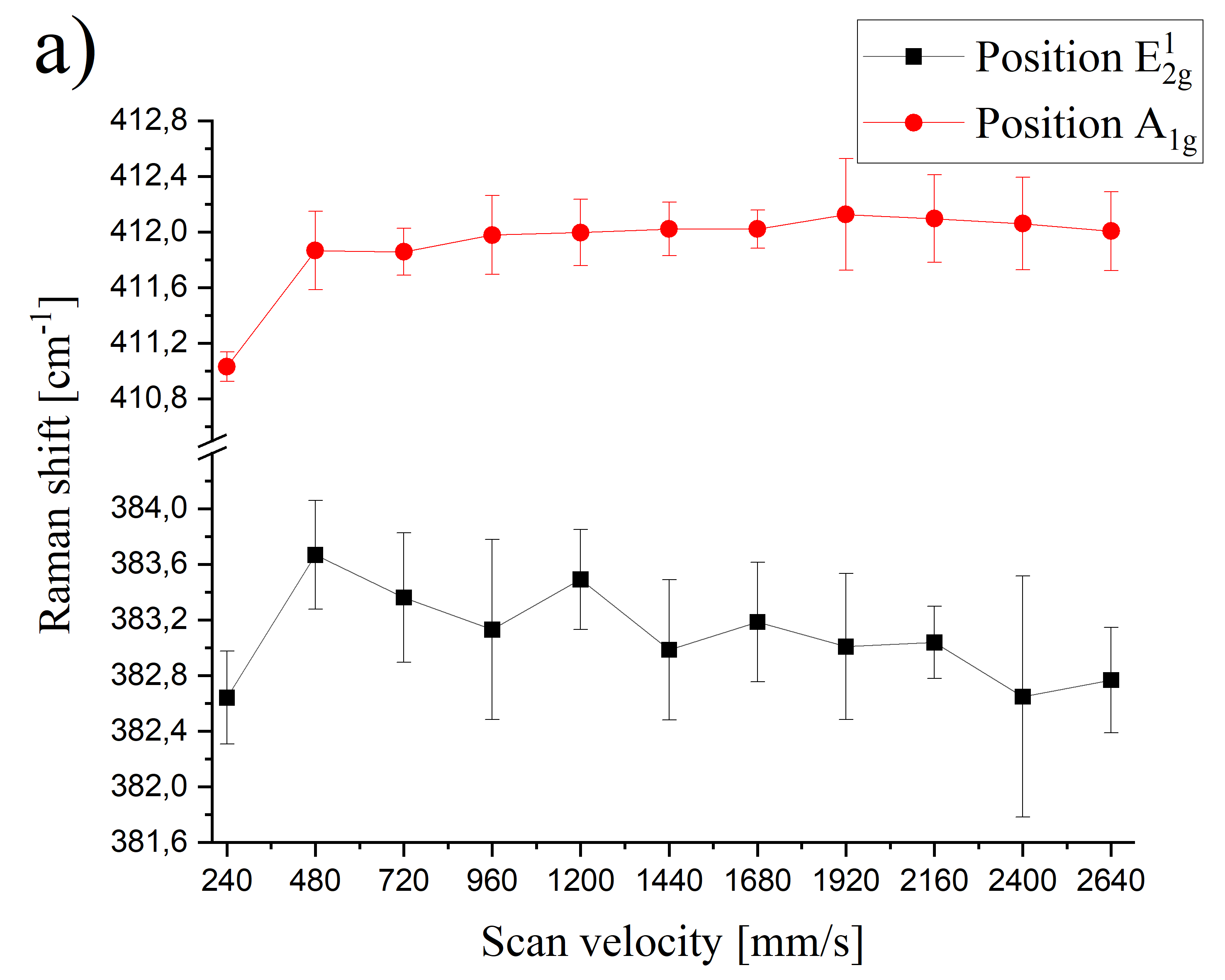}
			\includegraphics[width=0.49\linewidth]{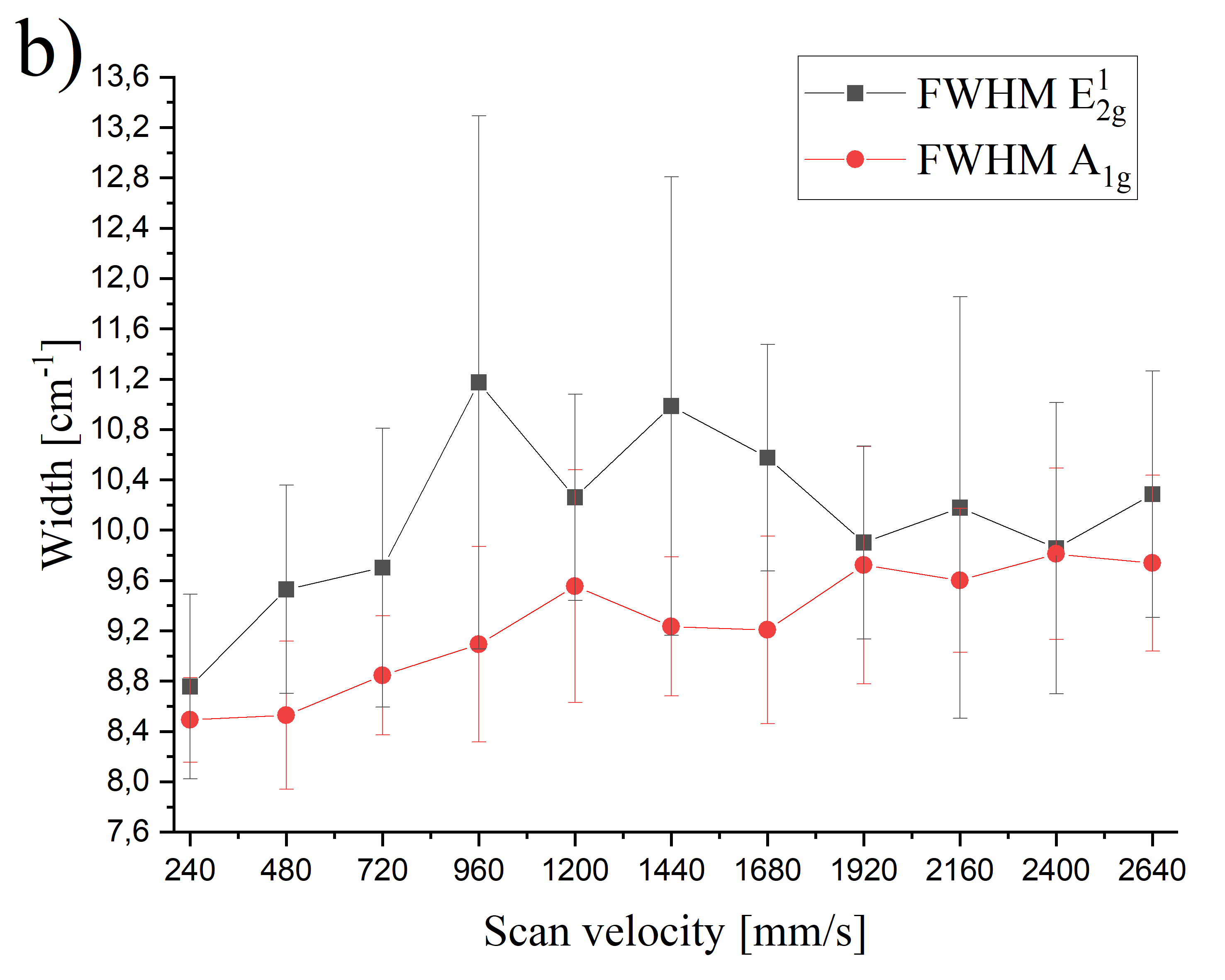}		
			\caption{a) Influence of the scanning speed (P\textsubscript{L} = 175 mW) on the position of the $\text{E}^1_{2g}$ -  and $\text{A}_{1g}$ - peak of the laser processed areas. b) Influence of the scanning speed (P\textsubscript{L} = 175 mW) on the FWHM of the $\text{E}^1_{2g}$ -  and $\text{A}_{1g}$ - peak of the laser processed areas.}\label{Auswertung_Geschwindigkeiten}
		\end{figure}
This redshift can be induced by an increasing number of defects \cite{Mignuzzi.2015,Parkin.2016} or biaxial tensile strain \cite{OBrien.2017} in the film. In comparison to the $\text{E}^1_{2g}$ - peak the position of the $\text{A}_{1g}$ - peak does not differ between the laser powers, but has a blueshift in comparison to the values stated in literature (literature: $\sim$ 407.8 cm\textsuperscript{-1} \cite{Li.2012}, laser annealed: $\sim$ 411.1 cm\textsuperscript{-1}). This blueshift can be associated with an out-of-plane growth of the MoS\textsubscript{2} films and the resulting curvature strain \cite{Virsek.2007,Krause.2009b,Berning.2022} or a doping of the film \cite{Chakraborty.2012}.
Besides the laser power, the scanning speed influences the annealing results. For a more detailed analysis the position and the FWHM of the $\text{E}^1_{2g}$ - and $\text{A}_{1g}$ - peak depending on the scanning speed with a laser power P\textsubscript{L} = 175 mW were extracted, shown in  Figure \ref{Auswertung_Geschwindigkeiten}(a) and (b). Regarding the position, both peaks shift to higher Raman shifts, when the scan velocity increases from 240 mm s$^{-1}$ to 480 mm s$^{-1}$. Afterwards the position of the $\text{A}_{1g}$ - peak remains stable, whereas at around 2400 mm s$^{-1}$ the $\text{E}^1_{2g}$ - peak shifts back to the starting position. In addition to the position, the FWHM’s of both peaks shift to larger numbers with increasing scanning speed, indicating a smaller crystal size and higher defect density. This behavior fits to the annealing process, since a slower scanning speed means exposure to a higher number of laser pulses. This could lead to a temperature accumulation, where the maximum temperature increases with the number of pulses \cite{Hoppius.2018,Kwon.2019}.

\begin{figure}
		\includegraphics[width=0.3\linewidth]{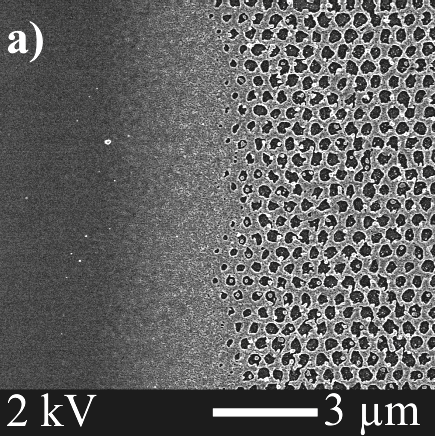}
	\hfill
	\includegraphics[width=0.3\linewidth]{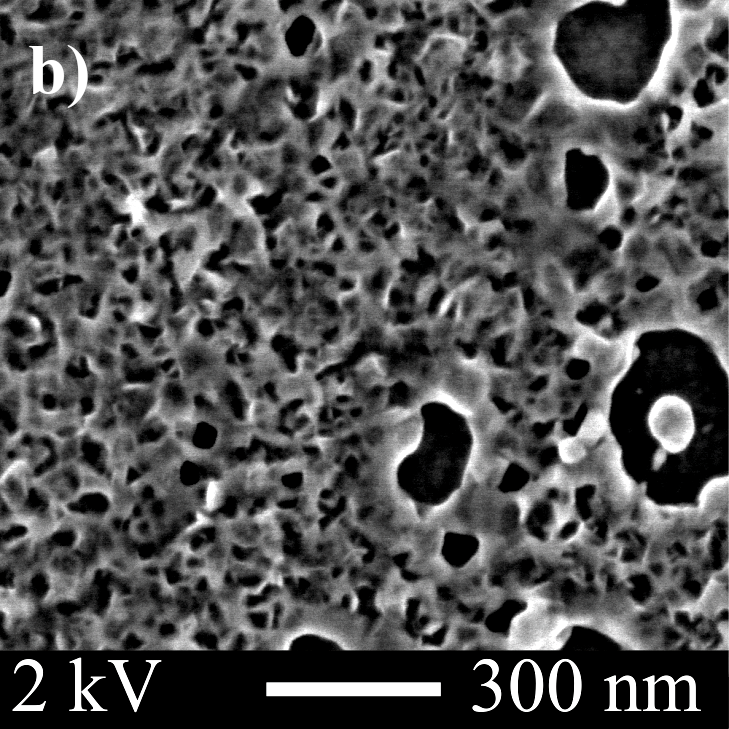}
	\hfill
	\includegraphics[width=0.3\linewidth]{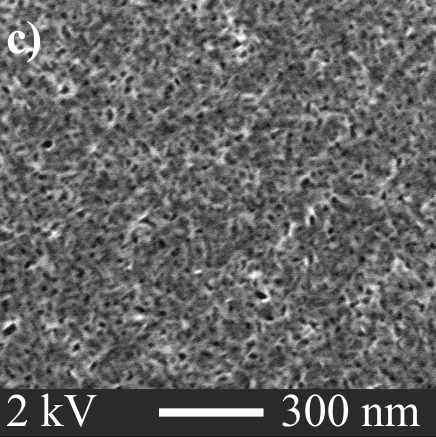}
	\caption{SEM images of usp laser processed areas. a) and b) is from the edge of a processed area with P\textsubscript{L} = 209 mW, v\textsubscript{scan} = 240 mm s$^{-1}$. c) is from the center of a processed area with P\textsubscript{L} = 175 mW, v\textsubscript{scan} = 240 mm s$^{-1}$. The images show the modification of the films, where a) and b) have a solidified pattern of the MoS\textsubscript{2} film where the underlying substrate gets revealed. In the solidified parts and the outlet of the processed area a disordered film gets visible. With lower laser powers the film surface is sealed, but the growth reminds disordered with out-of-plane textures.}\label{SEM_Bilder}
\end{figure}

For a better insight in the modification, SEM images of the laser-processed areas were taken. Figure \ref{SEM_Bilder}(a) displays the edge of the processed area with P\textsubscript{L}~=~209~ mW and v\textsubscript{scan}~=~240~mm s$^{-1}$. The smooth surface on the left image side is the unprocessed area, changing into a rougher surface at the edge of the processed film (center of the image). On the right side of the image, the film disrupts. The small holes (300 - 400 nm diameter) in the film are much smaller than the laser spot size (27 $\mu$m) and larger than the distance between two following laser shots ($\Delta x = v_{scan}/f_{rep} = 120\; nm$), indicating hydrodynamic flow in the molten MoS\textsubscript{2} layer. The hexagonal-like symmetry of the pattern is also typical for self-organization in hydrodynamic systems \cite{Fermigier.1992}, and can be observed also in laser-induced nanostructures \cite{Gurevich.2011,Chen.2009}. The resulting nano-pattern also explains the different color in the light microscopy image. In Figure \ref{Proben_uebersicht}(a) it can be observed, that at higher scanning speeds the nanopattern doees not start from the first line. This indicates line-to-line accumulation confirming that thermal processes like melting play an important role here. As we show later, the pulse-to-pulse temperature accumulation can be neglected. However, the defects generated by the first line can change the film properties and facilitate the pattern formation by the following lines. In Figure \ref{SEM_Bilder}(b) a higher resolution image of Figure \ref{SEM_Bilder}(a) is depicted, where the edge of the processed area is shown. The size of the nano-holes increases with the laser energy absorbed by the surface. The SEM image figure \ref{SEM_Bilder}(b) reveals that the processed MoS\textsubscript{2} crystallizes in small, disordered crystals. This explains the blueshift of the Raman peaks in comparison to thermally annealed samples. The effect of lower laser powers (P\textsubscript{L} = 175 mW) on the film modification is presented in Figure \ref{SEM_Bilder}(c). Here the SEM image was taken from the center of the processed area. The surface of the film has no nano-patterns, but appears rougher in comparison to the pristine film in Figure \ref{SEM_Bilder}(a). The SEM images explain the difference in the Raman spectra. Because of the melting of the film by the highest laser power, the lattice has more time to arrange itself and form larger crystals with tensile strain resulting in narrower Raman peaks and the shift of the $\text{E}^1_{2g}$ - peak. The shift of the $\text{A}_{1g}$ - peak is induced by the higher defect density and the curvature strain, that is induced by the out-of-plane growth during the laser annealing.

\subsection{Temperature estimation during usp laser process}
The mechanism of energy transfer from the laser to the film is not completely understood up to now. Principally, annealing is a temperature driven process, where the crystals need a specific temperature to arrange themselves \cite{Martin.1997}. During the usp laser processing different mechanisms can occur to heat the film. The required energy could be absorbed directly in the MoS\textsubscript{2} film converted to heat or exciting electrons. Or the substrate absorbs the energy, heats up and “bakes” the top laying film. A third possible effect is the occurrence of the “etalon effect”, leading to the interference between the incident light and the light reflected from the substrate forming an interference pattern reducing the amplitude of the electromagnetic wave in a thin (much smaller than $\lambda$) layer near to the surface. The influence of this effect on the laser ablation was studied by Ho et. al \cite{Ho.2014} and Solomon et al. \cite{Solomon.2022}. Solomon et al. suggested that thin film interference in the substrate can change the substrate reflectivity and therefore influence the ablation threshold of MoS\textsubscript{2}.  We neglect this effect in our estimation, because the angle-dependent change in the reflectivity of the pristine layers was not observed. First the absorbed energy and the absorption mechanism in the MoS\textsubscript{2} film and the SiO\textsubscript{2}/Si substrate were estimated. The pulse intensity at a power of P\textsubscript{L}~=~175~mW  and a focus diameter d\textsubscript{f}~$\approx$~30~$\mu$m is I\textsubscript{p}~$\approx$~1~GW cm\textsuperscript{-2}. For a 40 nm thick bulk-like MoS\textsubscript{2}, a transmittance T~$\approx$~0.32 is reported \cite{Ermolaev.2020}. Therefore $\sim$70 \% gets absorbed by the linear 1-photon absorption, whereas nonlinear absorption can be neglected, because of $\beta$~=~7.6~$\cdot$~10\textsuperscript{3}~cm GW$^{-1}$ \cite{Li.2015}.  With a reflectance of R~$\approx$~0.5 [33] the absorbed intensity is  I\textsubscript{abs}~$\approx$~0.34~GW cm\textsuperscript{-2}. In comparison, the transmittance of SiO\textsubscript{2} is T~$\approx$~0.99 \cite{RodriguezdeMarcos.2016}. Therefore, the remaining energy that passes through the MoS\textsubscript{2} film will only get absorbed by 1\% resulting in an intensity I\textsubscript{SiO\textsubscript{2}}~$\approx$~3.4~ MW cm\textsuperscript{-2} distributed over the whole oxide thickness. The rest is absorbed in the Si substrate, which has a good thermal conductivity and is separated from the MoS\textsubscript{2} by a thick SiO\textsubscript{2} film. The distribution of the absorbed intensities indicates that the baking effect of the substrate can be neglected and the direct absorption is the main mechanism for the heat development. The temperature $\theta$\textsubscript{l} reached after one laser pulse can be estimated with equation \ref{Temperatur_Abschätzung.}:
\begin{equation}
	E_p(1-r)=C_p m(\theta_l - \theta_0 )
	\label{Temperatur_Abschätzung.}
\end{equation}
 where the initial temperature $\theta$\textsubscript{0}~=~300~K we expect it to be much less than $\theta$\textsubscript{l} and neglect. The specific heat capacity is C\textsubscript{p}~$\approx$~400~J kg$^{-1}$ K$^{-1}$ \cite{Volovik.1978}. The mass $m$ of the heated material can be calculated with:
  \begin{equation}
  	m = \rho Ah
  \end{equation}
   where $h$ is the depth, at which the incident light is absorbed and $A$ is the focus spot area. As soon as $\sim$~70~\% of the incident light is absorbed in the  $l$~=~40~nm layer, we assume $h \approx l$. With a pulse energy of E\textsubscript{p}~=~90~nJ the surface temperature reaches up to $\theta$ $\sim$ 1000 K, which is not enough to melt the material ($\theta$\textsubscript{melt}~=~2600~K) \cite{Windholz.1983}.  Nevertheless, an oxidation of the films could occur at temperatures of $\theta$\textsubscript{oxid}~$>$~350~K \cite{Windom.2011}. As the Raman spectra of the annealed films do not show any MoO\textsubscript{x} peaks \newline ($\Delta\omega$~=~820~cm\textsuperscript{-1}) \cite{Windom.2011}, we assume, that the time interval, where the film temperature reaches the oxidation temperature, is not long enough to form MoO$_x$ in the film. The heat flow from the MoS\textsubscript{2} film into the underlying SiO\textsubscript{2} film is defined by the Thermal Boundary Condition (TBC) between the two materials. The TBC in literature differ between G~$\approx$~10~MW m\textsuperscript{-2} K$^{-1}$ - G~ $\approx$~33~MW m\textsuperscript{-2} K$^{-1}$ for MoS\textsubscript{2} on SiO\textsubscript{2} \cite{Suryavanshi.2019}. If we take the lowest TBC the cooling heat flow is $q \approx 7\; W$ calculated with equation \ref{Wärmefluss}.
   \begin{equation}
		q = GA\theta 
		\label{Wärmefluss}   
   \end{equation}
  with $\theta$~=~1000~K. From this follows, that the surface cools down after $t_C\approx 10\;ns$ calculated with:
  \begin{equation}
  	{t}_{C}=\frac{{E}_{A}}{q}\quad with \quad E_A=(1-r) E_p
  \end{equation}   
  In comparison to the time interval between two pulses $\delta=1/f_{rep}=500\;ns$, the surface cools down 50 times faster making heat accumulation unlikely. 
\subsection{Comparison of different annealing methods for MoS\textsubscript{2}}
For a comparison of the conventional annealing methods and  usp laser annealing, identical samples of the ALD produced MoS\textsubscript{2} films were processed by thermal and cw laser annealing. For the usp laser parameter set the spectra of the parameterset P\textsubscript{L} = 175 mW and v\textsubscript{scan} = 240 mm s$^{-1}$ were chosen. These parameters were selected, because they produced a continuous film (see Figure \ref{SEM_Bilder} c)) and had the smallest peak FWHM's (see Figure \ref{Proben_uebersicht} b) and Figure \ref{Auswertung_Geschwindigkeiten} b))  As seen in Table \ref{Tabelle_Annealing} after all the annealing methods the $\text{E}^1_{2g}$ and $\text{A}_{1g}$ - peaks appear in the Raman spectra, indicating a rise of the degree of crystallinity towards the amorphous films. The raw spectra of the samples can be found in Figure S1 in the supporting information. Because of the 40~nm thick films a bulk behavior of the films is expected. This behavior is represented in the thermally annealed films with the $\text{E}^1_{2g}$  - peak position around 384.9~cm\textsuperscript{-1} and the $\text{A}_{1g}$ - peak position around 408.2~cm\textsuperscript{-1}. In comparison to that, the peak position of the usp laser annealed films differs from the cw and thermal annealed peaks. Both laser methods induce a redshift of the $\text{E}^1_{2g}$ - peak ($\Delta\lambda_{cw}$~=~380.0~cm\textsuperscript{-1} and $\Delta\lambda_{usp}$~=~382.3~cm\textsuperscript{-1}) for the annealed films. An increase in defect quantity and a decrease in the grain size is indicated by the broadening of the FWHM of the $\text{E}^1_{2g}$ - peak and the $\text{A}_{1g}$ - peak \cite{Andrzejewski.2018,Mignuzzi.2015}. The redshift of the $\text{A}_{1g}$ - peak of the cw laser annealed sample can be explained by biaxial tensile strain \cite{OBrien.2017} or doping of the film \cite{Chakraborty.2012}. The doping could be oxygen related, occurring because of the processing of the films under natural atmosphere. In comparison to the cw laser annealed sample in the usp laser processed sample the $\text{A}_{1g}$ - peak induced a blueshift. This blueshift can be assigned to the out-of-plane grow of the MoS\textsubscript{2} films and the resulting curvature strain \cite{Virsek.2007,Volovik.1978}. But also a doping of the usp laser annealed MoS$_2$ film can not be excluded completely, because of the natural atmosphere during the processing. To get a more detailed insight, if the shift of the $\text{A}_{1g}$ - peak is doping or strain related, experiments are needed, which are carried out under inert gas or sulfur saturated atmosphere.  \\
\renewcommand{\arraystretch}{1.5}

\begin{figure}
	\centering
	\includegraphics[width=0.4\linewidth]{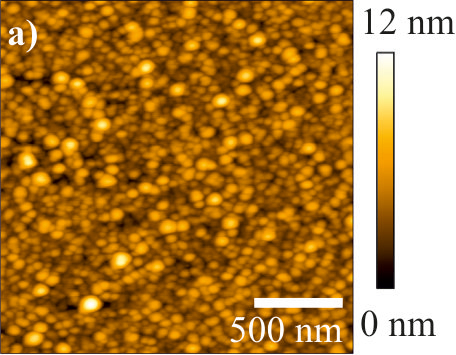}
	\includegraphics[width=0.4\linewidth]{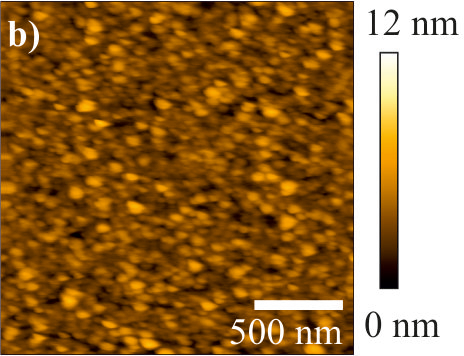}
	\\
	\includegraphics[width=0.4\linewidth]{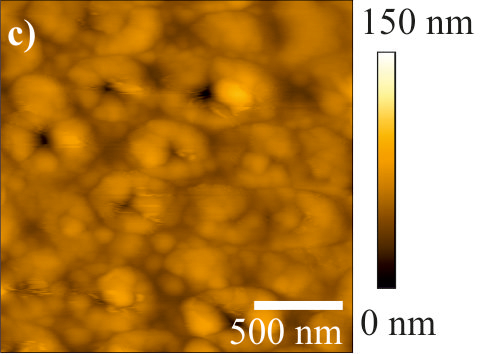}
	\includegraphics[width=0.4\linewidth]{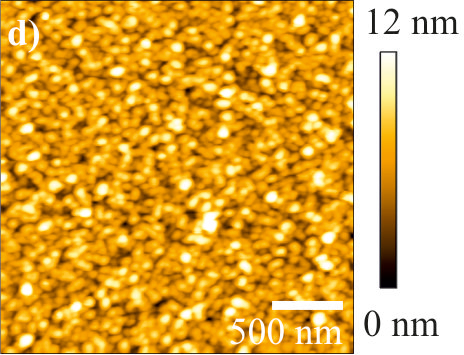}
	\caption{AFM images from the a) pristine, b) thermally annealed, c) cw laser annealed and d) usp laser annealed sample provides information about the surface structure and roughness. The surface structure produced by the annealing methods differ from each other. Thermal annealing produces the typical triangular structure of poly-crystalline MoS$_2$ films , whereas cw and usp laser annealing produces rougher chaotic structures. }\label{AFM_Bilder}
\end{figure}

	\begin{table}
			\caption{Fitted intensity ratio, position and FWHM of the $\text{E}^1_{2g}$ and  $\text{A}_{1g}$ - peaks from the Raman spectra of the different annealing methods. The values are in form: average ± standard deviation over six measurements. Ratio I\textsubscript{E}/I\textsubscript{A} is the ration of the maximum amplitudes of the $\text{E}^1_{2g}$ and $\text{A}_{1g}$ peaks.}
		\begin{tabular}[htbp]{@{}llll@{}}
			\hline
				& \textbf{thermal} & \textbf{cw} & \textbf{usp}\\
			\hline
		\textbf{Ratio I\textsubscript{E}/I\textsubscript{A}} & 0.37 ± 0.03 & 0.38 ± 0.03	
	& 0.36 ± 0.02 \\

	\textbf{Position $\text{E}^1_{2g}$ [cm\textsuperscript{-1}]}& 384.9 ± 0.2 & 380.0 ± 0.6 & 382.6 ± 0.3 \\

	\textbf{Position $\text{A}_{1g}$ [cm\textsuperscript{-1}]}& 408.2 ± 0.1 & 406.6 ± 0.7 & 411.0 ± 0.1 \\

	{\textbf{Position $\text{A}_{1g}$ - $\text{E}^1_{2g}$ [cm\textsuperscript{-1}]}} & 23.3 & 26.6 & 28.4 \\

	\textbf{FWHM $\text{E}^1_{2g}$ [cm\textsuperscript{-1}]} & 5.7 ± 0.9 & 7.6 ± 0.7 & 8.8 ± 0.7
	\\

	\textbf{FWHM $\text{A}_{1g}$[cm\textsuperscript{-1}]} & 6.8 ± 0.3 & 7.0 ± 0.6 & 8.5 ± 0.3
	\\
			\hline
		\end{tabular}
	\label{Tabelle_Annealing}
\end{table}
AFM images provide more information about the topography of the pristine and annealed MoS\textsubscript{2} films. Figure \ref{AFM_Bilder}(a) shows the pristine film, with no sign of triangular crystal growth, indicating the amorphous growth of the low-temperature ALD process \cite{Neubieser.2022}. After thermal annealing (Figure \ref{AFM_Bilder}(b)), the small poly-crystalline, triangular MoS\textsubscript{2} crystals have formed, indicating the successful process \cite{Tan.2014}. Because of the high film thickness the crystals appear to stack on top of each other, forming a continuous film from the small triangular patterns. Figure \ref{AFM_Bilder} (c) and (d) show AFM images of the laser annealed MoS$_2$ films. Both topographies do not exhibit triangular shapes, explaining the wider FWHM’s in the Raman spectra. In comparison to the pristine film, differences in the topography can be observed. The cw annealed MoS$_2$ film (Figure \ref{AFM_Bilder}(c)) shows large clusters of material that stacks on top of each other. The material clusters are so large that a 12 times higher height-scale has to be implemented. In SEM images, see Figure S2, dewetting of the film can be observed, with stacked material on the edges. The topography of the usp laser annealed MoS$_2$ film, see Figure \ref{AFM_Bilder}(d), reveals a smaller grain size in comparison to the cw laser annealed MoS$_2$ film. The topography appears similar to the thermally annealed topography, except that there are no triangular shaped particles and the maximum roughness is higher. The higher roughness (R\textsubscript{q\textsubscript{usp}} = 1.84 nm, R\textsubscript{q\textsubscript{thermal}} = 1.02 nm) can be explained with the disordered out-of-plane growth of the MoS\textsubscript{2} film, also seen in Figure \ref{SEM_Bilder} b) and c).

\section{Conclusion}
This work demonstrated that usp laser annealing is suitable for the annealing of amorphous MoS\textsubscript{2} films. The developed process shows the potential to be done on temperature-sensible substrates, without damaging them. With a suitable parameter setup an areal, sealed film can be produced, that exhibit a rise in the degree of crystallinity. A transition from homogeneous annealing to the formation of self-organized nano-patterns was observed, as the average laser power increased. Nevertheless, it should be noted that the usp processed films differ from the ‘classical’ annealed films in terms of crystal size, film stress and growth behavior. Future research is necessary to clarify the influence of these parameters on the electrical properties in the application. Furthermore, more research is necessary with respect to thinner film processing down to monolayer films and the question, if the process can be optimized regarding the film qualities and transferred on flexible substrates.  

\section{Experimental Section}
\subsection{Atomic layer deposition of MoS\textsubscript{2}}
The ALD of MoS\textsubscript{2} films was performed on a PICOSUN R-200 reactor with tetrakis(dimethylamido)-
molybdenum (TDMA-Mo) and H\textsubscript{2}S as precursor and co-reactant at 100 \degree C process temperature \cite{Neubieser.2022}. The deposition was carried out on 200 mm silicon wafers with 200 nm undoped silicate glass (USG), grown by PE-CVD. The number of growth cycles was 500, resulting in a film thickness of 40 nm.
\subsection{Annealing of amorphous MoS\textsubscript{2}}
\subsubsection{Ultrashort pulse laser annealing}
Ultrashort pulse laser annealing was performed with a HyperRapid NXT laser ($\lambda$ = 532 nm, t\textsubscript{p}~=~10~ps), with a repetition rate of f\textsubscript{rep} = 2 MHz. The laser power was adjusted by a half - wavelength ($\lambda$/2-wave plate) and a polarization beam splitter. Afterwards the laser beam is led to a galvanometer scanner, which is used to scan the target surface with the laser beam and focus the laser spot on the sample surface, with a spot size of d $\geq$ 27 $\mu$m. The galvanometer scanner covered rectangular areas by a set of parallel lines. Each line was shifted vertically by $\Delta y$ = 2 $\mu$m.
\subsubsection{Thermal annealing} 
Thermal annealed samples were prepared to compare with the usp laser-annealed MoS\textsubscript{2} layers. The annealing was performed in a furnace at 800 °C within a saturated sulfur atmosphere at 10 mbar. The sulfur atmosphere was achieved by placing elemental sulfur in front of the annealing zone and heating it to 130 \degree C. The annealing time was one hour if not stated otherwise.
\subsubsection{Cw laser annealing} 
Cw alloy experiments were performed using the Heidelberg DWL 66 FS laser lithography system equipped withe a 120 mW laser diode of wavelength 405 nm. For this purpose, a 2 mm write head was used without any further filters, and a maximum laser diode power of 96 mW was determined in the beam path.
\subsection{Characterization of the processed MoS\textsubscript{2} films}
Raman measurements were performed with a Renishaw InVia microscope ($\lambda$ = 532 nm) with a laser power P\textsubscript{L} $\leq$ 2 mW and an exposure time t\textsubscript{E} = 1 s to prevent the samples from heating up. Unless otherwise noted, all Raman spectra are averages of 6 measurements taken at different locations on the measured sample. For the evaluation of the Raman spectra the characteristic MoS\textsubscript{2} peaks were fitted with a Lorentzian fit. An example for a fitted Raman spectra can be seen in Figure S3.\\
The atomic force microscopy (AFM) measurements were conducted a surface area of 4 $\mu m^2$ using a Nanoscope Multimode V microscope from Digital Instruments, operating in tapping mode. The root-mean-square roughness was evaluated with NanoScope Analysis software.\\
SEM images were captured with a Schottky field emission electron microscope Leo Gemini 982 at 2 kV.

\medskip
\textbf{Supporting Information} \par 
Supporting Information is available from the Wiley Online Library or from the author.

\medskip
\textbf{Acknowledgements} \par 
The authors are grateful for funding and support from the Federal Ministry of Education and Research Germany (BMBF) BMBF-ForMikro-FlexTMDSense project, grant number 16ES1096K and 16ES1097. Further, the author thanks Pulsar Photonics GmbH for the provision of the usp laser system for the experiments. J.-L. W. thanks the Stiftung der deutschen Wirtschaft (sdw) within the Klaus Murmann fellowship for supporting his PhD project.

\medskip

%


%
%

\end{document}